\documentclass{article}
\usepackage{dcase2019_02}
\usepackage{amsmath}
\usepackage{graphicx}
\usepackage{url}
\usepackage{times}
\usepackage{booktabs}
\usepackage{tabularx}
\usepackage{color}
\usepackage{comment}
\usepackage{multirow}
\usepackage{setspace}

\newlength\savedwidth
\newcommand{\wcline}[1]{\noalign{\global\savedwidth\arrayrulewidth\global\arrayrulewidth 1.0pt} \cline{#1}
\noalign{\global\arrayrulewidth\savedwidth}}

\title{Overview of Tasks and Investigation of Subjective Evaluation Methods in Environmental Sound Synthesis and Conversion}

 \name{Yuki Okamoto$^{1}$,
       Keisuke Imoto$^{1}$,
       Tatsuya Komatsu$^{2}$, 
	Shinnosuke Takamichi$^{3}$,
       }
 \secondlinename{	  
       Takumi Yagyu$^{1}$,
       Ryosuke Yamanishi$^{1}$, 
       Yoichi Yamashita$^{1}$
       }
 \address{$^1$ Ritsumeikan University, Japan, $^2$ LINE Corporation, Japan, $^3$ The University of Tokyo, Japan\\
  }

\begin{document}
\ninept
\maketitle
\begin{sloppy}
\begin{abstract}
Synthesizing and converting environmental sounds have the potential for many applications such as supporting movie and game production, data augmentation for sound event detection and scene classification. Conventional works on synthesizing and converting environmental sounds are based on a physical modeling or concatenative approach. However, there are a limited number of works that have addressed environmental sound synthesis and conversion with statistical generative models; thus, this research area is not yet well organized. In this paper, we review problem definitions, applications, and evaluation methods of environmental sound synthesis and conversion. We then report on environmental sound synthesis using sound event labels, in which we focus on the current performance of statistical environmental sound synthesis and investigate how we should conduct subjective experiments on environmental sound synthesis.
\end{abstract}
%
\begin{keywords}
Environmental sound synthesis, environmental sound conversion, sound event synthesis, sound scene synthesis, subjective evaluation, WaveNet
\end{keywords}
%
\section{Introduction}
\label{sec:intro}
%
Sound synthesis and conversion are techniques for generating a natural sound using a statistical model that associates input information with the generated sound.
Sound synthesis and conversion methods with the aim of generating speech or music have been widely developed \cite{Zen_SC2009_01,Mohammadi_SC2017_01,Briot_arXiv2017_01}.
Recently, some researchers have also developed methods for environmental sound synthesis and conversion
that can be applied to support movie and game production \cite{Lloyd_ACMI3DGG_01}, the generation of content for virtual reality (VR) \cite{Zhou_CVPR2018_01}, and data augmentation for sound event detection and scene classification \cite{Salamon_WASPAA2017_01}.
Many studies on environmental sound synthesis and conversion have taken a physical modeling or concatenative approach \cite{Schwarz_DAFx2011_01,Bernardes_SMCC2016_01,Salamon_WASPAA2017_01}.
On the other hand, there have been fewer studies on environmental sound synthesis and conversion based on statistical generative models such as deep learning approaches.
To the best of our knowledge, there is no literature giving an overview of the problem definitions and evaluation methods for environmental sound synthesis and conversion.
Moreover, there have been no investigation of subjective evaluation methods for environmental sound synthesis and conversion.

In this paper, we therefore review problem definitions, applications, and evaluation methods of environmental sound synthesis and conversion.
We then report on environmental sound synthesis based on WaveNet \cite{Oord_CoRR2016_01}, which successfully synthesizes human voices, to discuss the current performance of statistical environmental sound synthesis.
Moreover, we investigate subjective evaluation methods of environmental sound synthesis.

The remainder of this paper is structured as follows.
In Sec. 2, we review problem definitions of environmental sound synthesis and conversion, their applications, and evaluation methods.
In Sec. 3, subjective experiments carried out to evaluate the performance of sound event synthesis using a WaveNet-based method are reported.
Finally, we summarize and conclude this paper in Sec. 4.
%
%
\begin{figure}[t!]
\centering
\includegraphics[scale=0.325]{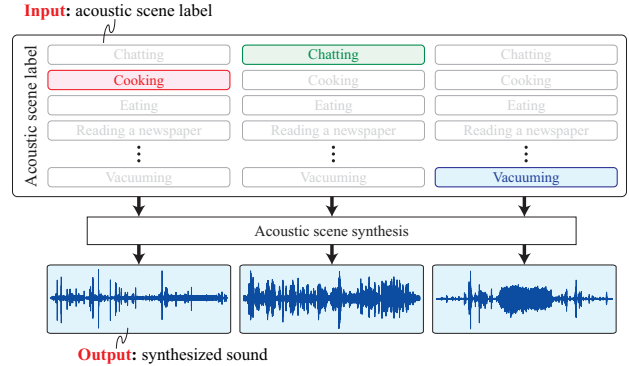}
\vspace{-5pt}
\caption{Problem definition of sound scene synthesis}
\label{fig:SSS}
\end{figure}
\vspace{2pt}
%
\begin{figure}[t!]
\centering
\includegraphics[scale=0.325]{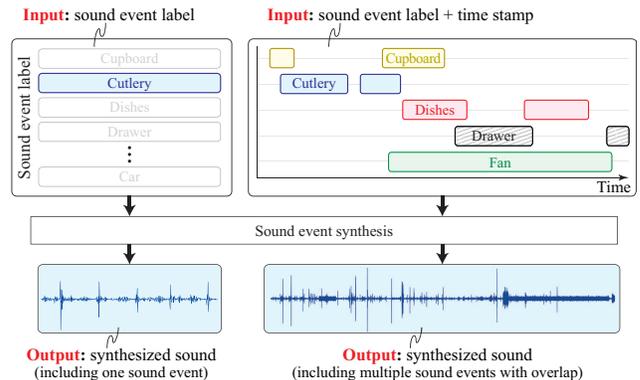}
\vspace{-13pt}
\caption{Problem definition of environmental sound synthesis using sound event labels}
\label{fig:SES01}
\end{figure}
%
%
%
\section{Problem Definitions of Environmental Sound Synthesis and Conversion}
\label{sec:problem}
In this section, we review applications, problem definitions, and evaluation methods of environmental sound synthesis and conversion, specifically environmental sound synthesis using event or scene labels (Sec. \ref{ssec:ESS01}), environmental sound synthesis using onomatopoeic words (Sec. \ref{ssec:ESS02}), environmental sound conversion (Sec. \ref{ssec:ESC}), and environmental sound synthesis/conversion using multimedia (Sec. \ref{ssec:multimodal}).
%
%
\subsection{Environmental Sound Synthesis Using Sound Event and Scene Labels}
\label{ssec:ESS01}
When providing movies or games with background sounds or sound effects, we need to listen to many sounds in a large sound database and select the most suitable one for the scene or sound event, which is a time-consuming part of movie or game production.
To address this issue, a statistical method for synthesizing an environmental sound well representing a sound event or scene, which utilizes the sound event or scene labels as below as an input, has been proposed \cite{Kong_ICASSP2019_01}.
Figures \ref{fig:SSS} and \ref{fig:SES01} illustrate the processes of environmental sound synthesis using the sound event or scene labels as the inputs of the systems, where we call these research tasks {\bf sound event synthesis (SES)} and {\bf sound scene synthesis (SSS)}, respectively.
Another issue is that the construction of an environmental sound dataset is very time-consuming compared with the construction of a speech or music dataset \cite{Imoto_AST2018_01}.
In recent studies, environmental sound analysis based on deep neural networks has required a large number of sounds to achieve a reasonable performance.
To overcome this problem of a shortage of environmental sound datasets, SES and SSS can be applied for data augmentation in environmental sound analysis.

To generate environmental sounds by a statistical approach, Kong \textit{et al.} \cite{Kong_ICASSP2019_01} have proposed a method of environmental sound synthesis utilizing a conditional SampleRNN \cite{Mehri_IClR2017_01} with sound scene labels represented as one-hot vectors.

A method of evaluating synthesized environmental sounds is an important subject in this research area.
When we apply SES or SSS to data augmentation for sound event detection or acoustic scene classification, it is reasonable to evaluate the methods of SES or SSS via their event detection or scene classification performance with augmented data.
On the other hand, in the case of utilizing the sound synthesized by SES or SSS itself, it has not been investigated in detail how the synthesis method should be evaluated.
In this paper, we focus on the subjective evaluation method for environmental sound synthesis in Sec. \ref{sec:experimets}.

%
\begin{figure}[t!]
\centering
\includegraphics[scale=0.33]{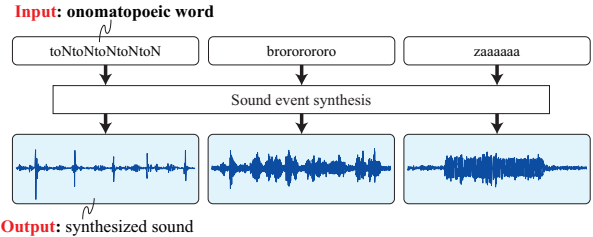}
\vspace{-5pt}
\caption{Problem definition of environmental sound synthesis using onomatopoeic words}
\label{fig:ESS02}
\end{figure}

On the other hand, the subjective evaluation of sounds is very time-consuming; thus, it is desirable to test methods for environmental sound synthesis and conversion with an objective evaluation of synthesized sounds.
There are some methods of objective evaluation such as the perceptual evaluation of speech quality (PESQ) \cite{PESQ_ITUT2001_01}, perceptual objective listening quality analysis (POLQA) \cite{POLQA_ITUT2011_01}, and perceived evaluation of audio quality (PEAQ) \cite{PEAQ_ITUT2001_01}, which are used for the evaluation of the speech quality in telecommunication or audio quality via codecs.
However, the objective evaluation method for synthesized or converted environmental sounds has not been investigated.
%
%
\subsection{Environmental Sound Synthesis Using Onomatopoeic Words}
\label{ssec:ESS02}
The SES and SSS discussed in Sec.~\ref{ssec:ESS01} control synthesized environmental sounds only using the sound event or scene labels; thus, they cannot control synthesized sounds without types of sound or scenes.
For instance, when synthesizing the sound of a car horn, it cannot be determined in advance whether SES will synthesize a horn sound with a continuous high tone (e.g, peeeeeeeeee) or one with an intermittent low tone (e.g, beep beep beep).
To control synthesized environmental sounds more finely, we can apply environmental sound synthesis using onomatopoeic words as an input of the system, as shown in Fig.~\ref{fig:ESS02}.
For SES using onomatopoeic words, Ikawa \textit{et al.} \cite{Ikawa_ICASSP2018_01} have proposed a method that converts onomatopoeic words to wave forms of environmental sounds using an encoder--decoder model.
%
%
%
\begin{figure}[t!]
\centering
\includegraphics[scale=0.325]{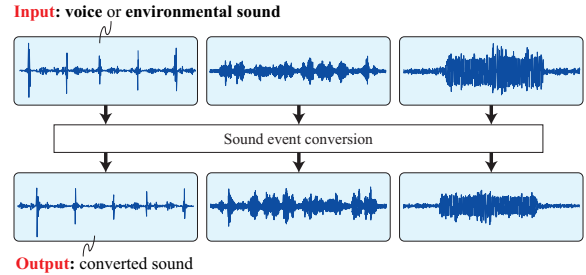}
\vspace{-5pt}
\caption{Problem definition of environmental sound conversion}
\label{fig:SEC}
\vspace{-6pt}
\end{figure}
\begin{table}[t]
\small
\caption{Experimental conditions}
\label{table:experimental}
\centering
\begin{tabular}{ll}
    \wcline{1-2}
    &\\[-8pt]
    Sound length & 1--2 s\\
    Sampling rate & 16000\\
    Wavefrom encoding & 16-bit linear PCM (real sound)\\
    & 8-bit $\mu$-law (synthesized sound)\\
    \cline{1-2}
    &\\[-8pt]
    Filter size & 2\\
    Learning rate & 0.001\\
    Batch size & 5\\
    Receptive field & 64 ms \\
    \# dilations & $2^{0}-2^{9}$\\
    \# residual channels & 32 \\
    \# dilation channels & 32 \\
    \# quantization channels & 256\\
    \# skip channels & 512\\
    \wcline{1-2}
\end{tabular}
%
\vspace{0pt}
%
\caption{Number of synthesized sounds used for subjective test}
\label{table:experimental_samples}
\centering
\small
\begin{tabular}{ccccc}
    \wcline{1-5}
    &\\[-8pt]
    \multirow{2}{*}{Experiment} & \multirow{2}{*}{\# labels} & \# samples & \multirow{2}{*}{\# listeners} & \# total\\[-1pt]
    & & in each label & & samples\\
    \wcline{1-5}
    &\\[-8pt]
    Exp. 1 & 10 & 5 & 24 & 1,200\\
    Exp. 2 & 10 & 4 & 24 & 960\\
    Exp. 3 & 10 & 2 & 24 & 480\\
    \wcline{1-5}
\end{tabular}
\end{table}
%
%
\subsection{Environmental Sound Conversion}
\label{ssec:ESC}
Sound event synthesis using onomatopoeic words is a flexible way of synthesizing environmental sounds; however, it is still difficult to control the generated environmental sounds as intended.
One way to address this problem is to synthesize environmental sounds not with sound event labels or onomatopoeic words but with the environmental sound or voice as the input of the system, as shown in Fig.~\ref{fig:SEC}.
We call this kind of task a {\bf sound event conversion (SEC)} or {\bf sound scene conversion (SSC)}.
When we have some background sounds or sound effects but they are not suitable for the movie or game, environmental sound conversion can also be applied to obtain desirable sounds.
For instance, when we have the horn sound of car X and a video including car Y, we can convert the horn sound of car X to that of car Y using SEC without re-recording the horn sound of car Y.

To convert environmental sounds to other audio signals, Grinstein \textit{et al.} \cite{Grinstein_ICASSP2018_01} and Mital \cite{Mital_arXiv2017_01} have applied a neural-style transfer-based method \cite{Gatys_CVPR2016_01}, which enables the ``style'' and ``content'' of an audio to be independently manipulated and copied to another audio signal.
\begin{figure}[t!]
\centering
\includegraphics[width=0.86\columnwidth]{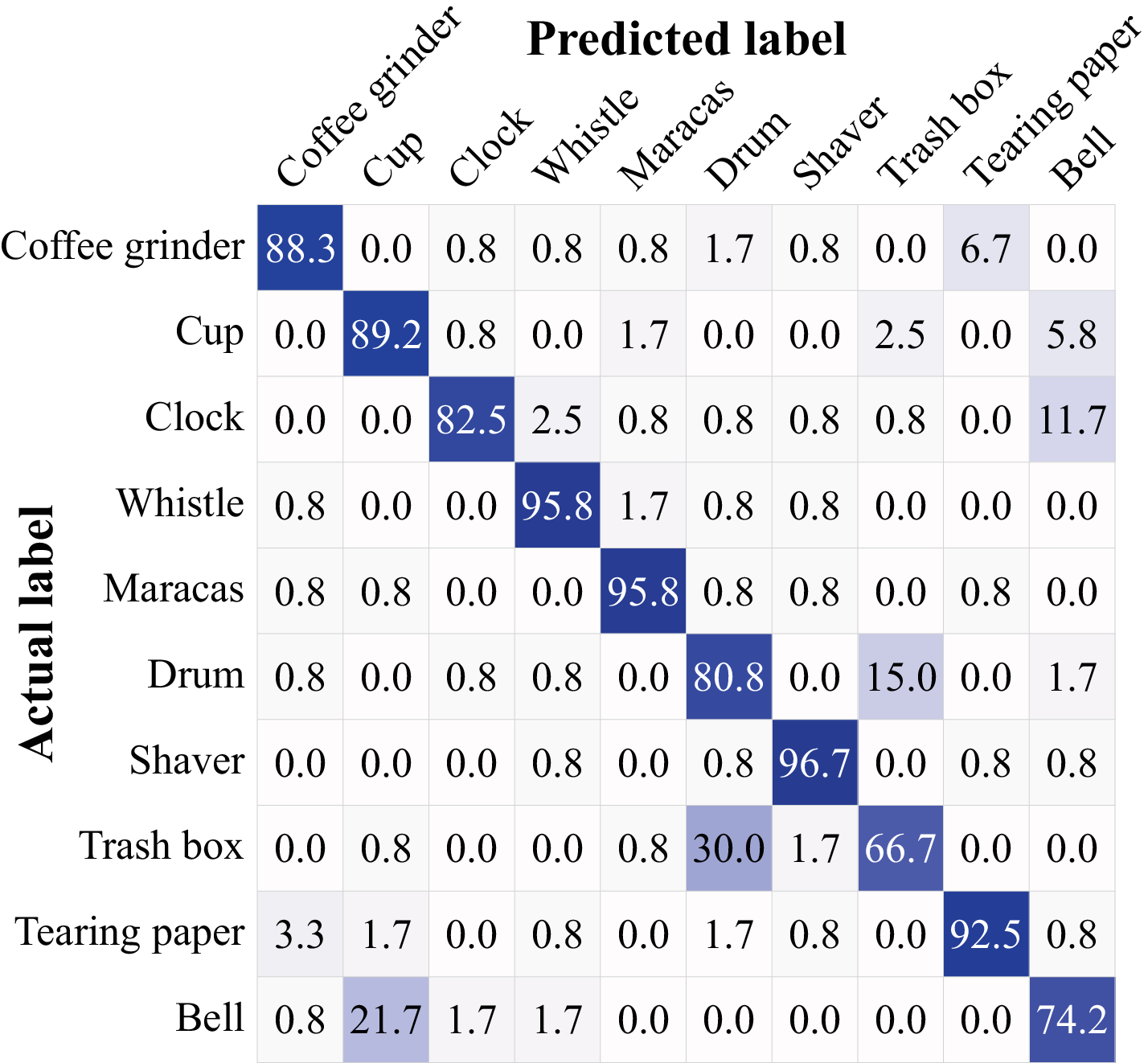}
\vspace{-3pt}
\caption{Confusion matrix of classification accuracy for original audio samples in terms of recall}
\label{fig:exp1_ori}
\end{figure}
%
%
%
\subsection{Environmental Sound Synthesis/Conversion Using Multimedia}
\label{ssec:multimodal}
Some researchers have addressed environmental sound synthesis and conversion using multimedia information as an input such as images.
For instance, Zhou \textit{et al.} have proposed a method for synthesizing environmental sounds from images that is based on SampleRNN \cite{Zhou_CVPR2018_01}.
%
%
%
\section{Investigation of Subjective Evaluation Method}
\label{sec:experimets}
\vspace{-3pt}
\subsection{Experimental Conditions}
\label{ssec:conditions}
\vspace{-3pt}
In this section, by evaluating SES using sound event labels based on the conditional WaveNet \cite{Oord_CoRR2016_01}, we discuss the current performance of environmental sound synthesis and how we should conduct a subjective test to evaluate a method under development.
For the evaluation, we considered 10 different sound events (manual coffee grinder, cup clinking, alarm clock ringing, whistle, maracas, drum, electric shaver, trash box banging, tearing paper, bell ringing) contained in the RWCP-SSD (Real World Computing Partnership-Sound Scene Database) \cite{Nakamura_LREC2000_01}.
We used a total of 1,000 samples (100 samples $\times$ 10 sound events), in which 95 samples of each sound event were used for model training and the others were used for the subjective test.
Table~\ref{table:experimental} shows the experimental conditions and parameters used for WaveNet.
Samples of sounds synthesized by WaveNet are available at \cite{Imoto_demos2019_01}.

\begin{figure}[t!]
\centering
\includegraphics[width=0.86\columnwidth]{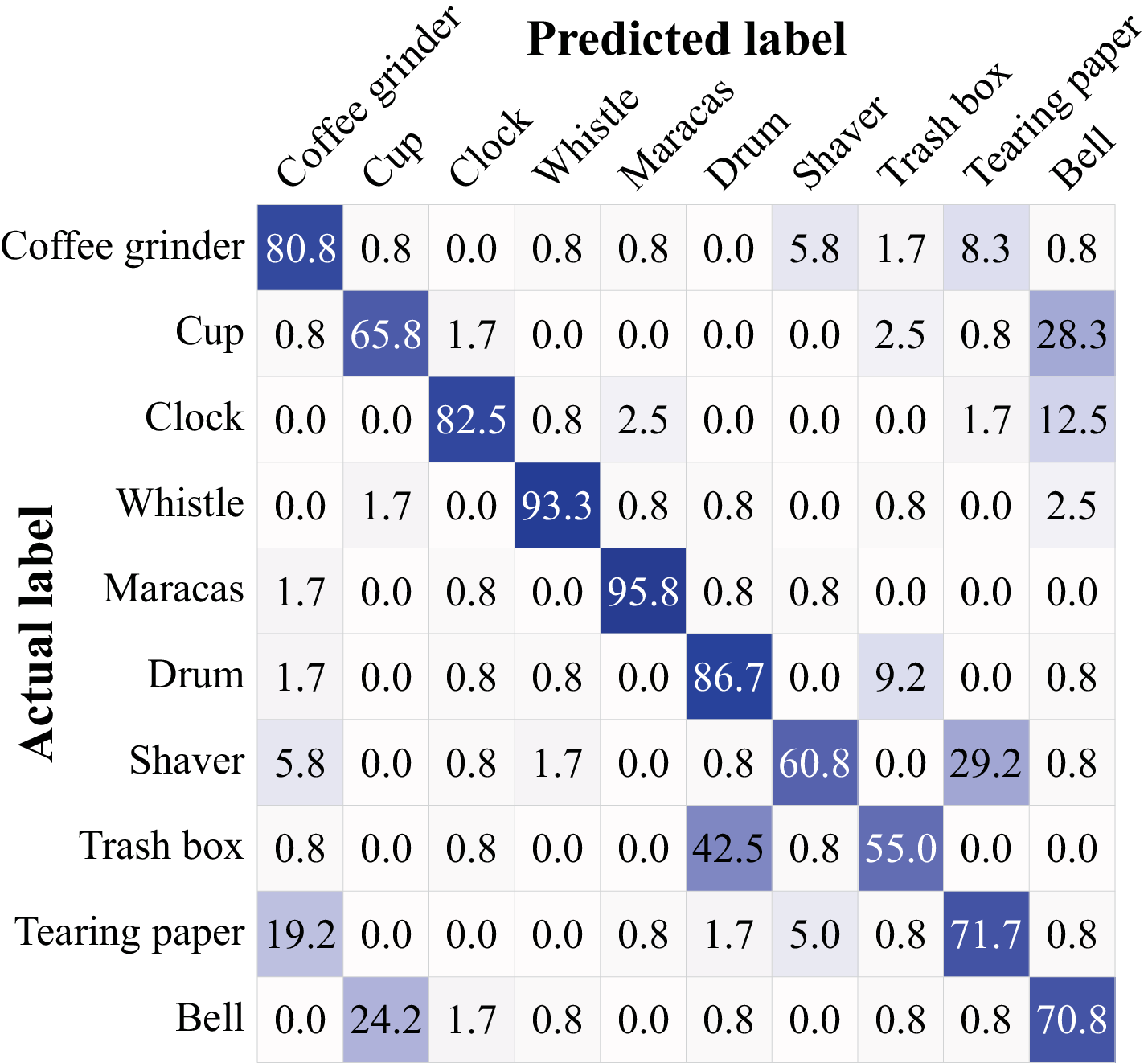}
\vspace{-3pt}
\caption{Confusion matrix of classification accuracy for synthesized audio samples in terms of recall}
\label{fig:exp1_gene}
\end{figure}
\begin{figure}[t!]
\centering
\includegraphics[width=1.0\columnwidth]{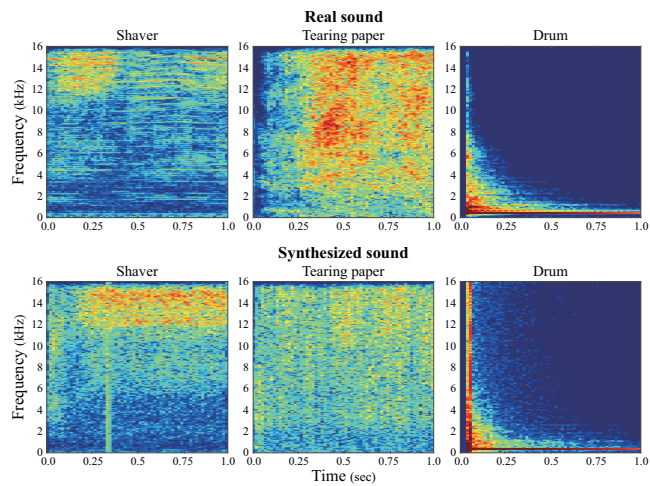}
\vspace{-17pt}
\caption{Spectrograms of real and synthesized environmental sounds}
\label{fig:spectrogram}
\end{figure}
%
%

Many works on speech and music synthesis have been conducted using subjective tests to evaluate the quality of synthesized sounds.
For example, in speech synthesis, speech intelligibility and naturalness are often used as evaluation metrics.
On the other hand, there have been no works in which methods of subjective tests in environmental sound synthesis and conversion were investigated in detail; thus, we here discuss how we should conduct subjective tests for environmental sound synthesis.
For synthesized sounds, it is important that (I) they are distinguishable from other types of environmental sound, (II) they are not distinguishable from real sounds, and (III) they have as high naturalness as real environmental sounds.
On the basis of these considerations, we conducted the following experiments:

\begin{itemize}
  \setlength{\itemsep}{-3pt}
  \item{\bf Experiment I: evaluation of intelligibility of synthesized sounds}\\
  After listening to a synthesized sound, the listener selected a sound event label that best represented the sound.
  As a comparison, the listener also similarly evaluated real environmental sounds.\\[-1pt]
  
  \item{\bf Experiment II: evaluation of distinguishability of real and synthesized sounds}\\
  We conducted a preference AB test.
  After listening to a pair of real and synthesized sounds in random order, the listener selected the one that sounded more real.\\[-1pt]
  
  \item{\bf Experiment III: evaluation of naturalness of synthesized sounds}\\
  We conducted a five-scale mean opinion score (MOS) test.
  After listening to a real or synthesized sound presented randomly, the listener scored the naturalness from 1 (very unnatural as an environmental sound) to 5 (very natural as an environmental sound). \\[-11pt]
\end{itemize}

Experiments were conducted with 24 listeners (13 males and 11 females) in a quiet environment at Ritsumeikan University.
In Table~\ref{table:experimental_samples}, the number of samples used in each experiment is listed.
In the experiments, a Roland QUAD-CAPTURE UA-55 audio interface and SONY MDR-CD900ST headphones were used.
%
%
%
%
\begin{figure}[t!]
\centering
\includegraphics[width=0.90\columnwidth]{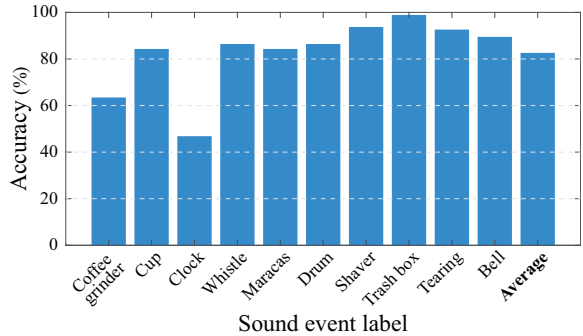}
\vspace{-6pt}
\caption{Recognition rate of real sounds}
\label{fig:exp2}
\end{figure}
%
%
\subsection{Experimental Results and Discussion}
\label{ssec:experiment}
{\bf Experiment I}: the classification results of real and synthesized sounds in terms of recall are shown in Figs.~\ref{fig:exp1_ori} and \ref{fig:exp1_gene}, respectively.
The averaged F-scores for real and synthesized sounds were 86.22\% and 76.30\%, respectively.
From these results, synthesized drum sounds are classified with a similar performance to real sounds, whereas the synthesized sounds of a cup clinking and an electric shaver tended to be more often misclassified than the real sounds.
Figure~\ref{fig:spectrogram} shows spectrograms of real and synthesized sounds.
This indicates that the synthesized sound of an electric shaver does not have the fine structure of the spectrum, which has the real sound.
Thus, the difference between the spectrograms of the sounds of an electric shaver and tearing paper is likely to be unclear, and this leads to the misclassfication.
From the results of experiment I, it considered that this subjective test is particularly helpful for evaluating whether the method can reproduce distinguishable sounds even when they have similar characteristics.

{\bf Experiment II}: listeners identified real sounds with an average accuracy of 82.71\% as shown in Fig.~\ref{fig:exp2}.
From this result, sounds synthesized by WaveNet do not have sufficiently high quality to be indistinguishable from real sounds.
This indicates that the evaluation of the distinguishability of real and synthesized sounds can be used for the comparison of conventional methods and more sophisticated methods of environmental sound synthesis that will be developed.

{\bf Experiment III}: the average MOS score for the naturalness of synthesized and real sounds and its 95\% confidence interval are shown in Fig.~\ref{fig:exp3}.
The results indicate that the synthesized sounds of the coffee grinder, clock, and maracas had similar naturalness scores to those of real sounds.
On the other hand, for the sounds of the electric shaver and the trash box banging, there are large differences in the MOS scores between the synthesized and real sounds.
We consider that this is because SES using WaveNet cannot reproduce the fine structure of the 
synthesized spectrum (e.g., the spectrum of the electric shaver in Fig.~\ref{fig:spectrogram}).
Moreover, Figs.~\ref{fig:exp1_ori}, \ref{fig:exp1_gene}, and \ref{fig:exp3} show that the listeners classified both the real and synthesized whistle sounds with reasonable performance, whereas there are large differences in the MOS scores between synthesized and real sounds.
This means that the evaluation of intelligibility is not satisfactory for evaluating the quality of synthesized sounds.

Thus, we propose that methods of environmental sound synthesis should be evaluated not only by testing the intelligibility of synthesized sounds but also by testing the distinguishability of real and synthesized sounds and/or the naturalness of synthesized sounds.
%
%
%
%
\begin{figure}[t!]
\centering
\includegraphics[width=0.88\columnwidth]{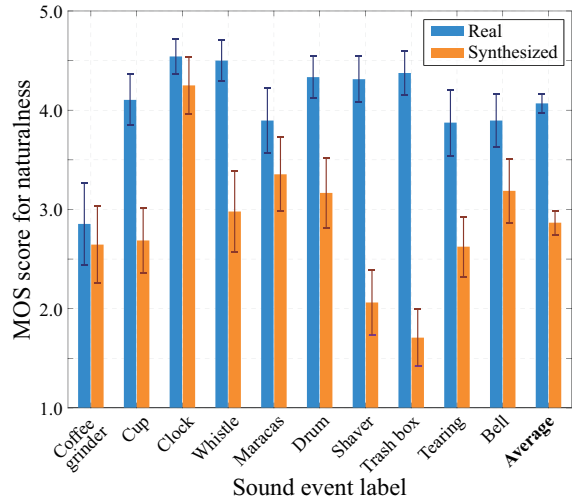}
\vspace{-5pt}
\caption{MOS score for naturalness of original and synthesized sounds}
\label{fig:exp3}
\end{figure}
%
%
\section{Conclusion}
\label{sec:conclusion}
In this paper, we presented the problem definitions of sound event synthesis, sound scene synthesis, and sound event and scene conversion.
We then discussed the current performance of sound event synthesis and subjective evaluation methods of environmental sound synthesis.
The evaluation experiments indicate that sounds synthesized by WaveNet do not yet have sufficiently high quality to be indistinguishable from real sounds.
Moreover, on the basis of our experimental results, we consider that methods of environmental sound synthesis should be evaluated by testing not only intelligibility but also distinguishability and/or naturalness.
%
%
%
%
\bibliographystyle{IEEEtran}
\bibliography{Okamoto_DCASE2019}
\end{sloppy}
\end{document}